\documentclass[aps,prd,11pt,twocolumn,superscriptaddress,
            floatfix,nofootinbib,longbibliography,reprint]{revtex4-2}

\usepackage{graphicx}  
\usepackage{bm}        
\usepackage{amssymb}   
\usepackage{amsmath}
\usepackage{mathrsfs}
\usepackage{relsize}
\usepackage[caption=false]{subfig}
\usepackage{mathtools}
\usepackage{xspace}
\usepackage{orcidlink}
\usepackage{soul}
\usepackage{enumitem}
\usepackage[final]{microtype}
\usepackage{multirow}
\usepackage{tabularx}
\usepackage{booktabs}

\usepackage{xcolor}
\usepackage{hyperref} 
\definecolor{dodgerblue}{HTML}{1E90FF}
\definecolor{DARK_ORANGE}{HTML}{FF5B00}
\definecolor{DARK_GREEN}{HTML}{02590F}
\definecolor{DARK_RED}{HTML}{B22222}
\definecolor{LIGHTER_DARK_RED}{HTML}{C86B6B}
\definecolor{dodgerblue}{HTML}{1E90FF}
\hypersetup{colorlinks=true, citecolor=DARK_RED,  urlcolor=dodgerblue, linkcolor=blue, pdfpagemode=FullScreen}

\newcommand{\orcidicon}[1]{\href{https://orcid.org/#1}{\includegraphics[height=\fontcharht\font`\B]{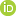}}}

\newcommand{\phXP}{\mbox{\textsc{IMRPhenomXP}}\xspace}

\newcommand{\phXPHM}{\mbox{\textsc{IMRPhenomXPHM}}\xspace}

\newcommand{\bilby}{\textsc{Bilby}\xspace}

\newcommand{\lalsuite}{\textsc{LALSuite}\xspace}

\newcommand{\dynesty}{\textsc{Dynesty}\xspace}
\newcommand{\bayeswave}{\textsc{BayesWave}\xspace}
\newcommand{\cwb}{\textsc{cWB}\xspace}

\newcommand{\numpy}{\mbox{\textsc{NumPy}}\xspace}
\newcommand{\scipy}{\mbox{\textsc{SciPy}}\xspace}
\newcommand{\matplotlib}{\mbox{\textsc{Matplotlib}}\xspace}

\newcommand{\Mgraviton}{\mbox{{\ensuremath{m_g = 7.07 \times 10^{-22}\:\mathrm{eV}/c^2}}}}

\newcommand{\Hz}{\ensuremath{\,\mathrm{Hz}}\xspace}

\newcommand{\alphamin}{\ensuremath{\alpha_{\mathrm{min}}}\xspace}
\newcommand{\alphamax}{\ensuremath{\alpha_{\mathrm{max}}}\xspace}

\newcommand{\forbital}{\ensuremath{f_{\mathrm{orbital}}}\xspace}
\newcommand{\forbitaltl}{\ensuremath{f_{\mathrm{orbital}}(t,\vec{\lambda})}\xspace}

\newcommand{\falphatl}{\ensuremath{f_{\alpha}(t,\vec{\lambda})}\xspace}

\newcommand{\sigmaS}{\ensuremath{\Lambda_S}\xspace}

\newcommand{\DSY}{\ensuremath{\mathcal{D}_S^Y}\xspace}

\newcommand{\msun}{\ensuremath{\,\mathrm{M}_\odot}\xspace}

\newcommand\avg[1]{\langle #1 \rangle}

\newcommand{\FDU}{Center for Astronomy and Astrophysics, \mbox{Center for Field Theory and Particle Physics}, and Department of Physics, Fudan University, Shanghai 200438, China.\vspace*{3pt}}

\newcommand{\UCLouvain}{Centre for Cosmology, Particle Physics and Phenomenology - CP3, Universit{\'e} Catholique de Louvain, Louvain-La-Neuve, B-1348, Belgium.\vspace*{3pt}}

\newcommand{\ROB}{Royal Observatory of Belgium, Avenue Circulaire, 3, 1180 Uccle, Belgium.\vspace*{3pt}}

\newcommand{\IITGn}{Department of Physics, Indian Institute of Technology Gandhinagar, Gujarat 382055, India.\vspace*{3pt}}

\newcommand{\NewUU}{School of Natural Sciences and Humanities, New Uzbekistan University, Tashkent 100000, Uzbekistan.}

\begin{document}
\title{Probing missing physics from inspiralling compact binaries via time-frequency tracks}

\author{Debtroy Das~\orcidlink{0009-0005-8097-7923}}
\affiliation{\FDU}

\author{Soumen Roy~\orcidlink{0000-0003-2147-5411}}
\affiliation{\UCLouvain}
\affiliation{\ROB}

\author{Anand S. Sengupta~\orcidlink{0000-0002-3212-0475}}
\affiliation{\IITGn}

\author{Cosimo Bambi~\orcidlink{0000-0002-3180-9502}\vspace*{3pt}}
\affiliation{\FDU}
\affiliation{\NewUU}

\date{\today}

\begin{abstract}
The orbital evolution of compact binary systems depends on the component masses and spins as well as the underlying theory of gravity. Because the orbital dynamics directly determine the frequency evolution of gravitational-wave signals, tracking this evolution provides a sensitive probe of general relativity (GR). We introduce a method that coherently stacks time-frequency pixel energies along the observed orbital-frequency trajectory while allowing controlled shifts along the frequency axis. The resulting stacked energy reveals a clear signature of the dominant quadrupole mode in GR. When an alternative theory of gravity is injected and analyzed with GR waveforms, the corresponding energy distribution in the time–frequency plane is markedly different. We use this behavior to formulate a new consistency test based on a normalized deviation between the observed signal and the posterior predictions from theoretical waveforms. Using waveforms in beyond-GR scenarios within the sensitivity of second-generation interferometers, we demonstrate the ability of this method to identify departures from GR. We also examine missing physics within GR itself by analyzing an eccentric binary and recovery with a quasi-circular model. Finally, we apply the method to GW190814 and find signatures consistent with higher-order multipoles.
\end{abstract}

\maketitle

\section{\label{sec:level1}Introduction}
 
The direct detection of gravitational waves (GWs) from compact binary coalescences~\cite{LIGOScientific:2016aoc, LIGOScientific:2018mvr, LIGOScientific:2020ibl, KAGRA:2021vkt} by Advanced LIGO~\cite{LIGOScientific:2014pky} and Advanced Virgo~\cite{VIRGO:2014yos} has opened a new laboratory for tests of General relativity (GR) in the strong and dynamic regime over the past ten years~\cite{LIGOScientific:2016lio, LIGOScientific:2018dkp, LIGOScientific:2019fpa, LIGOScientific:2020tif, LIGOScientific:2021sio}. The approaches for testing GR can be broadly categorized into theory-specific and theory-agnostic tests \cite{Bambi:2017khi}. The former approaches involve analyzing GW signals through the lens of both GR and alternative gravity theories, aiming to identify either a model that better interprets the data or constrain beyond GR parameters that are predicted in the alternative theory. Theories beyond GR often suggest modifications to the gravitational field of BHs, the generation of GWs, and the propagation of signals \cite{Yunes:2009hc, Kleihaus:2011tg,Becker:2021ivq, Giddings:2017jts, Tu:2023xab, Liebling:2012fv}. The theory-agnostic tests focus on identifying potential deviations from predictions based solely on GR. This prompted the development of theory-agnostic tests, where one introduces deviation in the GW signal phasing coefficients.

Since waveform models in alternative theories of gravity are not yet sufficiently developed to enable rigorous tests, and the true underlying theory may still be unknown, several null tests have been proposed to detect deviations from the predictions of GR. Although these null tests offer valuable ways to probe deviations from GR, most of them rely on parameterized modifications in the phase evolution of GWs, either during the inspiral phase~\cite{Yunes:2009ke, Arun:2006yw, Agathos:2013upa, Roy:2025gzv, Mehta:2022pcn, Krishnendu:2017shb, Puecher:2022sfm, Mahapatra:2023hqq, Mahapatra:2023ydi} or in the merger-ringdown phase~\cite{Brito:2018rfr, Maggio:2022hre, Roy:2025gzv, Carullo:2019flw, Isi:2019aib, Mahapatra:2023htd}. The statistical bounds on the parametrized deviations from GR depend on the assumptions and systematic errors of GW modeling \cite{Moore:2021eok, Hu:2022bji, Pang:2018hjb, Narayan:2023vhm, Saini:2023rto, Chandramouli:2024vhw}.

On the other hand, the importance of signal assumption is low for consistency tests, such as the residual analysis. This method investigates whether any signal remains after subtracting the best-fit GR waveform from the data~\cite{LIGOScientific:2016lio}. We search for coherent residuals across detectors using the \bayeswave framework~\cite{Cornish:2014kda} by employing a set of Morlet wavelets. Due to its broad parameter space of wavelet model, the method has significant flexibility in identifying coherent residual features. However, its sensitivity to capture deviations from GR is reduced unless those deviations are loud.

Another complementary method for testing consistency is evaluating the agreement between the reconstructed signal and the posterior waveform samples obtained from parameter estimation from model-based Bayesian inference~\cite{Ghonge:2020suv}. The reconstructed signal is typically obtained using the unmodeled frameworks such as \cwb~\cite{Klimenko:2005xv, Klimenko:2015ypf} and \bayeswave~\cite{Cornish:2014kda}, which reconstruct any coherent features in the data across the detector network. Recently, a semi-modeled approach has been introduced to reconstruct the signal~\cite{Roy:2022teu}. Instead of searching for coherent features, it uses the posterior waveforms to identify essential wavelets that can represent the signal more efficiently. These methods are sensitive to detecting deviations from GR or missing physics in the theoretical waveform model. However, the accuracy of the reconstruction strongly depends on the chirp mass, as the overlap at a fixed signal-to-noise ratio decreases with a lower chirp mass~\cite{Roy:2022teu}.

As the dynamics of compact binaries become highly relativistic during the late inspiral, their orbital evolution serves as a pristine laboratory for testing GR and searching for signatures of new physics. The fidelity with which current waveform models reproduce the true inspiral dynamics depends on the assumption that GR is the correct theory of gravity. If this assumption is violated, the inferred orbital evolution will deviate from the GR prediction~\cite{Alexander:2018qzg}. For example, dipole radiation in modified gravity theories causes binaries to inspiral more rapidly and merge sooner than in GR~\cite{Gerard:2001fm}. Even within GR, additional physical effects—such as the orbital hang-up effect~\cite{Campanelli:2006uy, Varma:2018rcg}, orbital eccentricity, spin precession~\cite{Apostolatos:1994mx, Hannam:2013oca}, or the presence of an astrophysical environment around the binary~\cite{Barausse:2014pra, Roy:2024rhe, CanevaSantoro:2023aol}—can modify the orbital evolution of BBHs. Consequently, the energy distribution of the GW signal on the time–frequency plane is expected to display systematic differences with respect to our baseline GR model. Similar systematic differences in the observed signal may also arise from modified GW propagation relative to that predicted by GR.

\begin{figure}[t]
	\centering
		\includegraphics[width=\linewidth]{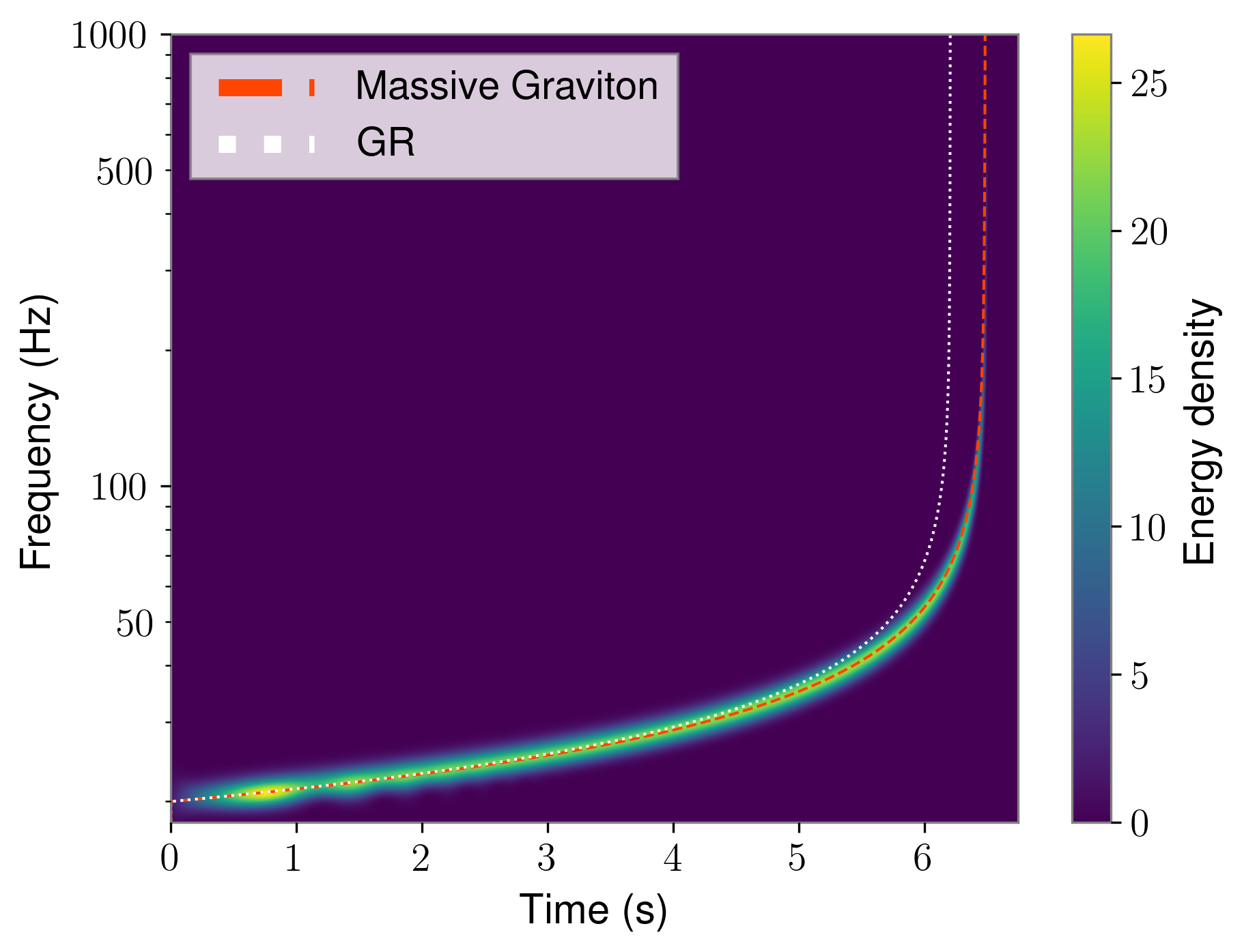} 
	\caption{Time-frequency representation of a GW170608-like massive graviton signal with a graviton mass of \Mgraviton, weighted by the Advanced LIGO designed sensitivity \texttt{aLIGOZeroDetHighPower}~\cite{aLIGO_ZDHP}. The red dashed  curve shows the time-frequency evolution of the massive graviton signal, while the white dotted curve shows for GR obtained assuming massless graviton.}
	\label{fig:track}
\end{figure}

In Fig.~\ref{fig:track}, we show the time–frequency representation of a massive-graviton signal and compare it to the GR expectation. The modified dispersion relation alters the propagation of GWs, producing a visibly different time–frequency evolution in the observed data. When such a signal is analyzed with GR templates, Bayesian inference attempts to match the track by shifting the inferred source parameters, thereby introducing systematic biases. While the GR waveform may align well with the dominant part of the signal track where most of the energy resides, but portions of the track will be systematically mismatched because GR cannot fully reproduce the modified propagation effects.

In this work, we develop a new consistency test by employing the time-frequency trajectory estimated using GR waveforms. We check whether the distribution of time-frequency pixel energies is consistent between the theoretical waveform and the observed signal. Instead of examining the pixels one-by-one, we sum the pixel energies along the estimated time-frequency trajectory and then perform the comparison. Since the true trajectory could follow a different curve, we slide our estimated trajectory up and down in frequency. For each small step, we sum the pixel energies along the corresponding time-frequency trajectory and then perform the comparison. Finally, we combine the comparison values over all small steps. This strategy allows us to collect any deviation from our theoretical waveform across the entire time-frequency spectrum.

Our proposed consistency test is motivated by previous work that used the TFR of GW signals from compact binary mergers to identify higher–order multipoles during the inspiral phase~\cite{Roy:2019phx}. This approach has been extensively employed by the LIGO-Virgo-KAGRA (LVK) Collaboration to detect higher–multipole radiation in GW190412~\cite{LIGOScientific:2020stg} and later found strong evidence for such modes in GW190814~\cite{LIGOScientific:2020zkf}. That method was specifically developed for identifying higher–order modes, where one examines the signal power only along the higher–mode tracks. Instead, to construct our new consistency test, we extend this idea to the entire time–frequency spectrum, comparing the distribution of signal power between the theoretical waveform and the observed data.

To demonstrate the performance of our method, we first consider a GW170608-like massive-graviton signal and compare it against our standard GR waveform that treats the graviton as massless. We then consider a numerical-relativity waveform from an eccentric binary, while the search template assumed to be a quasi-circular binary. Finally, we apply our method to real data from the GW190814 event to detect the presence of higher-order modes.

As discussed earlier, existing consistency-test methods either search for coherent residual power across a detector network or compute overlaps between reconstructed signals and best-fit posterior waveforms obtained through Bayesian inference. These approaches become less sensitive for detecting deviations as the signal duration increases. Instead, our time-frequency-trajectory-based consistency test remains sensitive even for longer signals.

The manuscript is organized as follows. Section~\ref{sec:level2} outlines the methodology of our consistency test. We provide a brief review of the literature that motivated this work and summarize the data-analysis and mathematical tools used to formulate the test. We then develop the statistical framework to evaluate the level of disagreement between the theoretical waveform and observed signal. In Section~\ref{sec:level3}, we demonstrate the validity of our method through a case study, followed by an application to GW190814 in Section~\ref{sec:level4}. Finally, Section~\ref{sec:level5} provides our conclusions and a discussion.

\section{Methodological Framework}
\label{sec:level2}

During the inspiral phase of a binary system, the instantaneous frequency $f_{\ell m}(t,\vec{\lambda})$ of a spherical harmonic mode $(\ell, m)$ is approximately related to the orbital frequency \forbitaltl as $f_{\ell m}(t,\vec{\lambda}) \approx m \forbitaltl$, where $\vec{\lambda}$ represents the set of binary parameters. The dominant gravitational-wave mode corresponds to $(\ell, m) = (2, 2)$, and the prominent “hot pixels” in the TFR trace this same frequency evolution.

To extract the energy distribution from the TFR, we define an arbitrary frequency track by scaling the orbital frequency, following Ref.~\cite{Roy:2019phx}:
\begin{equation}
\label{eq:ftrack}
    \falphatl = \alpha  \; \forbitaltl,
\end{equation} 
where $\alpha$ is a positive scaling factor. We scan $\alpha$ over a discrete range $[\alpha_{\rm min}, \alpha_{\rm max}]$. For each value of $\alpha$, we compute the energy along the corresponding frequency track by summing time-frequency pixels along the \falphatl:
\begin{equation}
\label{eq:stack}
    S(\alpha) = \mathlarger{\mathlarger{\sum}}_{t_{\mathrm{ISCO}} - \Delta t}^{t_{\mathrm{ISCO}}} \left \lvert \tilde{X}(t,f = f_{\alpha}(t,\vec{\lambda}) ) \right \rvert^2,
\end{equation}
where $\tilde{X}(t,f)$ is the TFR of the GW signal obtained by performing continuous wavelet transformation. The quantity $t_{\mathrm{ISCO}}$ is the time at which the orbiting masses reach the innermost stable circular orbit (ISCO), and  $\Delta t$ is the time-length of the $f_\alpha (t,\vec{\lambda})$ curve, which remains unchanged under scaling. 

This construction naturally allows us to probe harmonic content beyond the dominant quadrupole. For example, choosing $\alpha=2$ recovers the energy along the quadrupole $(m=2)$ track, while $\alpha=3$ corresponds to the octupole $(m=3)$, and so on. This strategy was originally introduced in Ref.~\cite{Roy:2019phx} to detect higher-order modes in GW signals. In that earlier work, the scaling was applied directly to the $(2,2)$ mode track, so the $\alpha$ values used there were half the values used in our current analysis.

We note that the quantity \forbital represents the observed orbital-frequency evolution rather than the true source-frame orbital dynamics. In the massive-graviton case, the intrinsic binary dynamics in the source frame are not modified; however, the altered propagation of gravitational waves causes the observed time–frequency evolution to deviate from the GR prediction, as illustrated in Fig.~\ref{fig:track}.

Throughout this work, we employ the synchroextracting transform~\cite{yu2017synchroextracting, pham2017high} to obtain a high-resolution TFR. By “high resolution,” we refer to improved localization of the signal energy in the time–frequency plane compared to continuous wavelet transformation. The mathematical formulation of synchroextracting transform is provided in Appendix~\ref{sec:A}, where we also show its advantages in our new consistency test.

\subsection{Setup for consistency test}
\label{sec:level2b}

In our study, we consider two types of injections: a beyond-GR signal assuming a massive graviton, and an eccentric waveform obtained from numerical simulations, as discussed in Sec.~\ref{sec:level3}. The massive graviton injection is generated for a GW170608-like system with aligned spins, assuming a graviton mass of \Mgraviton. This value exceeds the upper bound reported for the GW170608 event by LVK~\cite{LIGOScientific:2019fpa}, ensuring observable deviations from general relativity. A phenomenological modification to the GW dispersion relation is introduced by adding a power-law term in momentum to the GR dispersion relation $E^2 = p^2 c^2$, where $E$ and $p$ denote the energy and momentum of the GW, and $c$ is the speed of light. The modified dispersion takes the form \mbox{$E^2 = p^2 c^2 + A_\gamma p^\gamma c^\gamma$}, where $A_\gamma$ and $\gamma$ are phenomenological parameters. The exponent $\gamma$ can take values from 0 to 4 in increments of 0.5. The graviton mass is related to the $A_0$ case by $m_g = A_0^{1/2}/c^2$, where $A_0 > 0$ corresponds to a massive graviton, i.e., the dispersion relation for a massive particle in vacuum~\cite{Will:1997bb}. The waveform is generated using a phenomenological model that incorporates the effects of this dispersion into the GW phase evolution in the frequency domain. The correction to the phase, $\Phi(f)$, is given by $\delta\Phi(f) = - \pi c D_L/\lambda_g^2 f$, where $\lambda_g = h/(m_g c)$ is the Compton wavelength of the graviton, $h$ is Planck’s constant, and $D_L$ is the binary’s luminosity distance. We henceforth refer to this waveform as the “non-GR signal” and use it throughout to illustrate our methodology.

\begin{figure}[t]
	\centering
		\includegraphics[width=\linewidth]{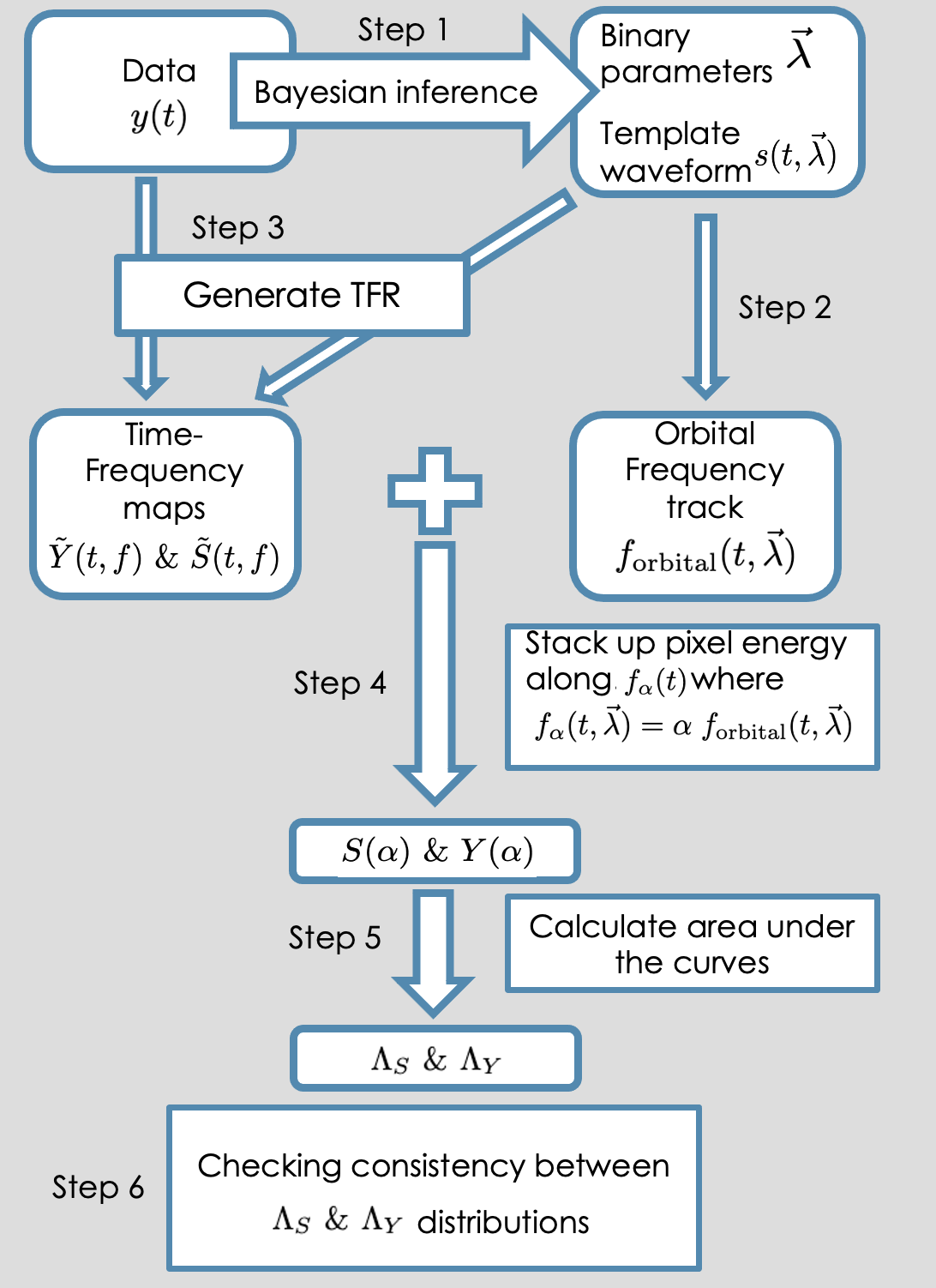} 
    \captionsetup{labelformat=empty}
	\caption{Flowchart summarizing the steps of our new time-frequency trajectory based consistency test for the GW signals from compact binary mergers.}
	\label{fig:flowchart}
\end{figure}

\begin{figure*}
	\centering
		\includegraphics[width=0.975\linewidth]{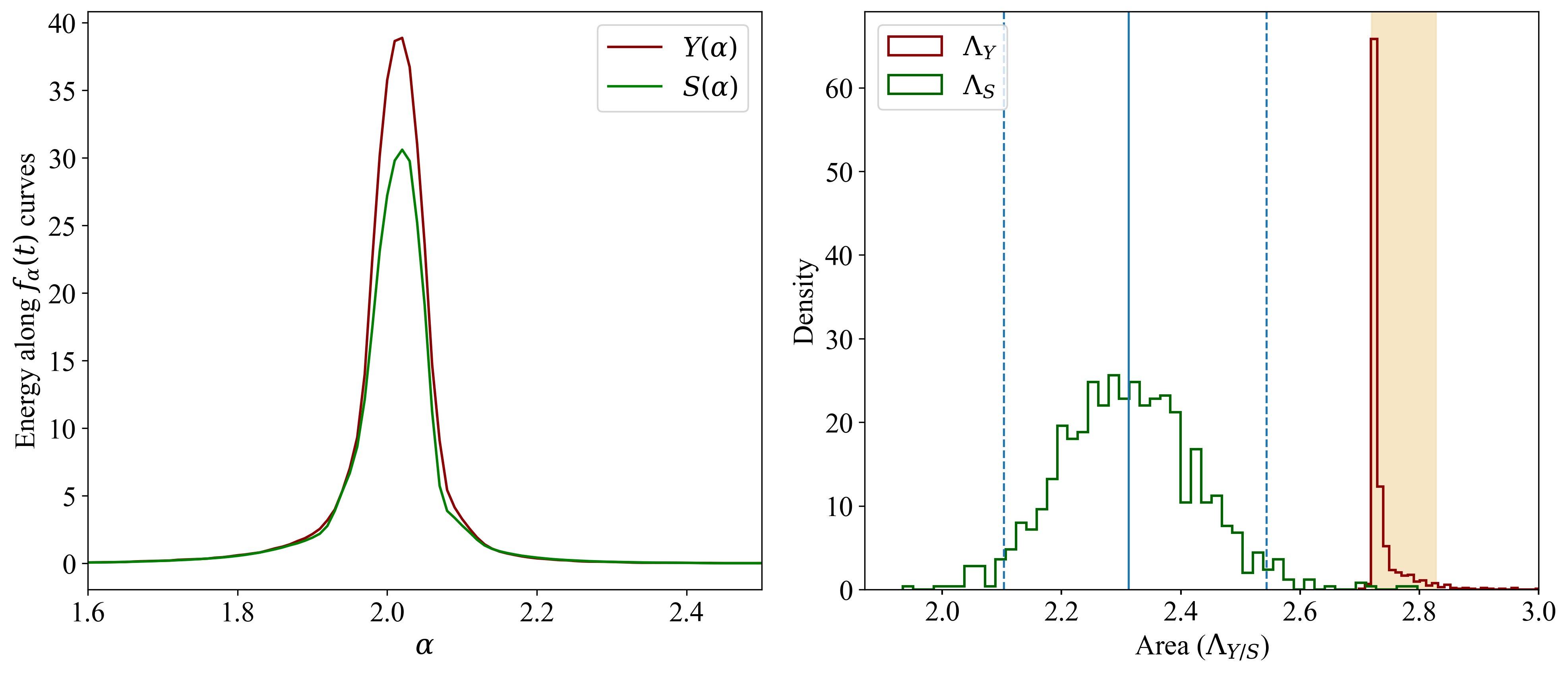} 
	\caption{\emph{Results for GW170608-like massive graviton injection and recovery with GR waveform}. The left panel shows the disagreement between the injections and best-fit template waveform. The $Y(\alpha)$ and $S(\alpha)$ curves are obtained using the injection signal and the template waveforms, respectively. The right panel shows the distributions for the template waveforms \sigmaS (in green) and the injected signal $\Lambda_Y$ (in maroon), as defined in Eq.~\eqref{eq:area}. The light brown band represents 95\% intervals of the $\Lambda_Y$ distribution. The blue dashed line denotes 95\% intervals of the \sigmaS distribution and solid blue line denotes its mean.}
	\label{fig:SalphaYalpha}
\end{figure*}

Throughout this work, we perform the injection studies assuming a network of three detectors -- Hanford (H1), Livingston (L1), and Virgo (V1) -- assuming their advanced designed sensitivities: \texttt{aLIGOZeroDetHighPower} for H1 and L1~\cite{aLIGO_ZDHP}, and \texttt{AdvVirgo} for V1~\cite{2012arXiv1202.4031M}, as implemented in \lalsuite~\cite{lalsuite}. We perform injection analyses assuming a starting frequency of 20\:\Hz and the network SNR is set to $\sim 43.5$.

We will now describe our method in the steps outlined below. This method is also summarized by the flowchart in Fig.~\ref{fig:flowchart}.

\begin{enumerate}[leftmargin=5mm]
    \item[] \emph{Step 1}: We perform Bayesian parameter estimation of the GW signal $y(t)$ (here the non-GR signal) using the \bilby~\cite{Ashton:2018jfp} package with \dynesty nested-sampling algorithm~\cite{2020MNRAS.493.3132S}. We use 1000 live points, sample method is \texttt{acceptance-walk}, and an acceptance rate of \texttt{naccept=60}. Throughout this work, we perform the analyses using the \phXP waveform model \cite{Pratten:2020ceb}, which is a phenomenological approximant designed to describe the full inspiral-merger-ringdown (IMR) evolution of BBH coalescences incorporating the orbital precession effects. Within the framework of GR, \phXP provides an efficient and accurate representation of GW signals from precessing binaries assuming quasi-circular orbits. For the analyses, we adopt uniform priors on the chirp mass and mass ratio, and a power-law prior for the luminosity distance. For the remaining parameters, we use the priors listed in Table 1 of Ref.~\cite{Ashton:2018jfp}.
    
    \item[] \emph{Step 2}: For each posterior sample, we compute the observed orbital frequency $\forbital(t,\vec{\lambda})$, where $\vec{\lambda}$ denotes the set of binary parameters associated with that sample.

    \item[] \emph{Step 3}: We whiten the data $y(t)$ and template waveform $s(t,\vec{\lambda})$ using the detector's power spectral densities. 
We then compute their time–frequency representations, denoted as $\tilde{Y}(t,f)$ for the data and $\tilde{S}(t,f)$ for the waveform.

    \item[] \emph{Step 4}: As discussed earlier, we generate the arbitrary frequency tracks $f_\alpha (t,\Vec{\lambda})$ using Eq.~\eqref{eq:ftrack}. This allows us to gradually scroll the $\forbital(t,\Vec{\lambda})$ curve along the frequency direction in the TF plane by changing the value of $\alpha$ in small discrete steps between $\alpha_{\rm min}$ and $\alpha_{\rm max}$. We stack up the pixel energy of $\tilde{Y}(t,f)$ using Eq.~\eqref{eq:stack} and obtain $Y(\alpha)$. Similarly, we obtain $S(\alpha)$ using $\tilde{S}(t,f)$. Any disagreement between the data (here, the non-GR signal) $y(t)$ and the template waveform $s(t,\Vec{\lambda})$ will propagate into the $S(\alpha)$ and $Y(\alpha)$ curves. The left panel of Fig.~\ref{fig:SalphaYalpha} shows the dissimilarity between the $S(\alpha)$ and $Y(\alpha)$ curves due to the massive graviton effect in the signal.

    \item[] \emph{Step 5}: The energy of the GW signals is well localized in both the $S(\alpha)$ and the $Y(\alpha)$ curves around $\alpha \thickapprox 2$. We calculate the area under these curves around the peak corresponding to the quadrupole mode in the range $\alpha \in [1.6,2.4]$ to quantify the mismatch in shape between $S(\alpha)$ and $Y(\alpha)$. The area under the $Y(\alpha)$ curve is defined as,
    \begin{equation} \label{eq:area}
    \Lambda_Y = \int_{\alphamin}^{\alphamax} Y(\alpha) \: d \alpha .
    \end{equation}
    We use the same equation to calculate \sigmaS (area under the $S(\alpha)$ curve) by replacing $Y(\alpha)$ with $S(\alpha)$. We compute $\Lambda_{Y/S}$ using the above equation for a single detector. For a network of detectors, we obtain the combined value $\Lambda_{Y/S}$ by adding the individual values from each detector.

    \item[] \emph{Step 6}: We randomly choose a large number of posterior samples. Then, for each posterior sample, we repeat Step 2 to Step 5 to construct the \sigmaS and $\Lambda_Y$ distributions. Throughout this work, we use 1000 posterior samples for each case study. The distributions of $\Lambda_Y$ and $\Lambda_S$ are shown in Fig.~\ref{fig:SalphaYalpha} (right panel) along with their 95\% intervals. If the theoretical waveform is consistent with the observed signal, we expect these two distributions to have significant overlap.

To quantify possible departures between the observed signal and our search waveforms, we compute the \emph{normalized deviation}, \DSY, of the value obtained from the data relative to the distribution obtained from posterior waveforms,
\begin{equation}
\label{eq:distance}
\DSY = \frac{\left|\Lambda_Y - \left<\Lambda_S\right>\right|}{\sigma_S},
\end{equation}
where $\left<\Lambda_S\right>$ and $\sigma_S$ denote the mean and standard deviation of $\Lambda_S$, respectively.  This definition is analogous in spirit to a z-score, i.e., the distance from the mean expressed in units of the standard deviation. The $1 \sigma$ ($\sim 68$\%) width of the $\Lambda_S$ distribution serves as the unit of distance measurement. We note that the average distance statistic $\langle \DSY\rangle$ indicates how many standard deviations the data lie away from the posterior prediction, providing a measure of the statistical significance of any deviations.

For the massive graviton injection, we find that \mbox{$\left<\DSY\right> = 3.75$}, which implies the $\Lambda_Y$ distribution lies, on average, outside the 99.999\% confidence interval of $\Lambda_S$ distribution. In Fig.~\ref{fig:SalphaYalpha} (right panel) the visual evidence that the 2 distributions are well separated ($95 \%$ intervals of the distributions do not overlap) is also justified by the distance calculation.

\end{enumerate}

\begin{figure}
	\centering
		\includegraphics[width=\linewidth]{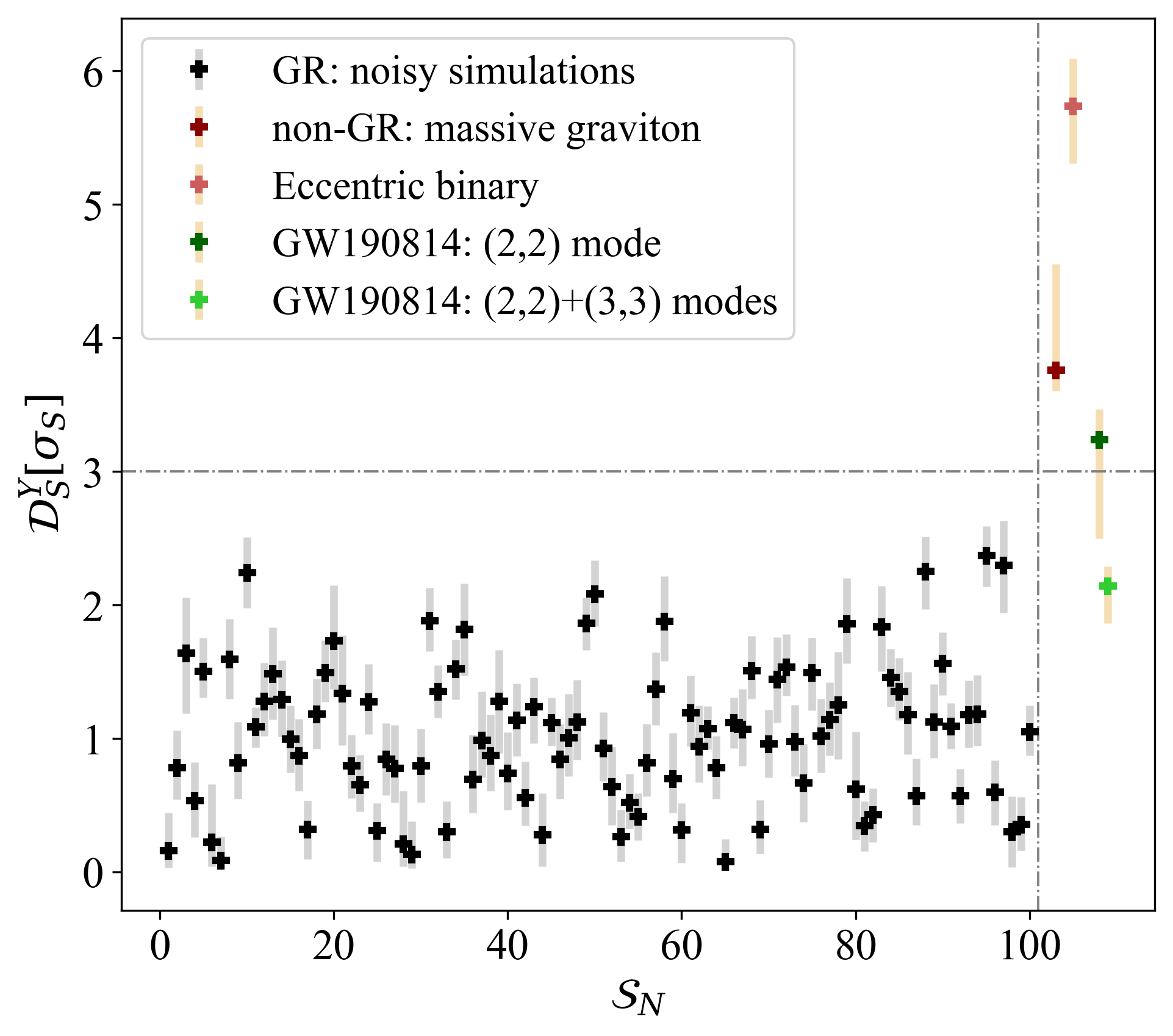} 
	\caption{\emph{Background for a known model with noise versus signals containing additional physics.} We show the distance statistic \DSY between the data and the search waveform model for three types of injection studies and for the GW190814 event. The GR injection study, used to assess the impact of noise in our framework where the injection and search waveforms are identical, is shown in black. The massive-graviton injection recovered with a GR model is shown in maroon, and the eccentric BBH injection recovered with a quasi-circular model is shown in orange. For GW190814, results are shown for two recovery models: one using only the dominant quadrupole mode (dark green) and another including higher-order modes (light green). The average of each \DSY distribution is marked with a “\textbf{+}”, and the shaded vertical bands denote the corresponding 95\% credible intervals.}
	\label{fig:d}
\end{figure}

\subsection{Can noise itself mimic deviation?}
\label{sec:level2c}

To evaluate the sensitivity of our framework, we perform an extensive injection study using simulated signals in noise, assuming that the injection and recovery waveforms are generated within the same model. The astrophysical GW signals observed by the LVK interferometers are embedded in detector noise; to mimic this scenario, we generate many independent noise realizations drawn from stationary Gaussian distributions weighted by the Advanced LIGO and Virgo sensitivity curves, as described in Sec.~\ref{sec:level2b}. We then inject an identical GW signal, generated using the \phXP model, into each noise realization. To quantify how noise alone may impact our proposed analysis, we repeat the full analysis pipeline on all noisy injections following the steps outlined in Fig.~\ref{fig:flowchart}. Using the posterior samples, we obtain the distributions for $\Lambda_Y$ and $\Lambda_S$ to evaluate the level consistency between data versus search waveform.

We demonstrate the distributions data versus search waveform in Fig.~\ref{fig:n} and Fig.~\ref{fig:n2} in Appendix~\ref{sec:B}. The summary of the distance statistic \DSY for 100 simulations is shown in Fig.~\ref{fig:d}. The range of \DSY for each simulation is represented by grey error bars, and the averages are marked with black plus symbols. All injections found below  $2.4$ when the true signal model is the same as the search waveform. We therefore set $\DSY = 3$ as a reference threshold in our framework for assessing whether disagreement between the search waveform and the observed data is driven by noise or by missing physics, as indicated by the horizontal dashed line.

In Fig.~\ref{fig:d}, we further highlight the significance of the massive graviton injection recovered with the standard GR waveform. We find a value of $\DSY \sim 3.8$, which unambiguously indicates that the disagreement cannot be explained by noise fluctuations alone, but instead points to genuine missing physics in the search waveform model.

\begin{table}[t!]
\centering
\setlength{\tabcolsep}{1mm}
\begin{tabularx}{0.48\textwidth}{l c}
\toprule[1pt]
\toprule[1pt]
Injection type & Average distance $\avg{\DSY}$ \\
\midrule[0.7pt]
GR injections with & $\left[0.08, 2.37\right]$ \\
many noise realizations &  \vspace{1mm} \\
Massive graviton & 3.76 \vspace{1mm} \\
Eccentric BBH & 5.84 \vspace{1mm} \\
GW190814: (2,2) mode & 3.24 \vspace{1mm} \\
GW190814: (2,2)+(3,3) modes & 2.14 \\ 
\bottomrule[1pt]
\bottomrule[1pt]
\end{tabularx}
\caption{\emph{Summary of the results from our proposed consistency test.} We report the average value of the distance statistic $\avg{\DSY}$, which quantifies the agreement between the data and the search waveform model, for three injection cases as well as for the GW190814 event. The GR-injection study uses identical injection and recovery waveforms to assess the impact of noise on our test. We also consider a massive-graviton injection recovered with GR waveforms, and an eccentric-BBH injection recovered with a quasi-circular model. The corresponding $\avg{\DSY}$ values are indicated with a `\textbf{+}’ symbol in Fig.~\ref{fig:d}.}
\label{tab:table1}
\end{table}

\section{Consistency test with signals from eccentric binary}
\label{sec:level3}

Not only can missing physics arise from beyond-GR effects, but our theoretical waveform may also lack important physics even within GR, such as unmodeled orbital eccentricity. Current LVK analyses predominantly assume quasi-circular binaries~\cite{LIGOScientific:2025slb}, motivated by the expectation that gravitational radiation efficiently circularizes most systems before they enter the sensitive band of ground-based interferometers. However, several astrophysical formation channels predict that binaries may retain measurable residual eccentricity at small separations. This has motivated substantial ongoing effort to develop eccentric waveform models~\cite{Gamboa:2024hli, Planas:2025feq, Albanesi:2025txj} and to search for evidence of eccentricity in LVK events~\cite{Romero-Shaw:2022xko, Gupte:2024jfe, Planas:2025jny}. Existing eccentric waveform families, though, are often computationally expensive or do not yet provide complete inspiral–merger–ringdown coverage with generic precession.

\begin{figure}[htbp]
    \centering
	\includegraphics[width=\linewidth]{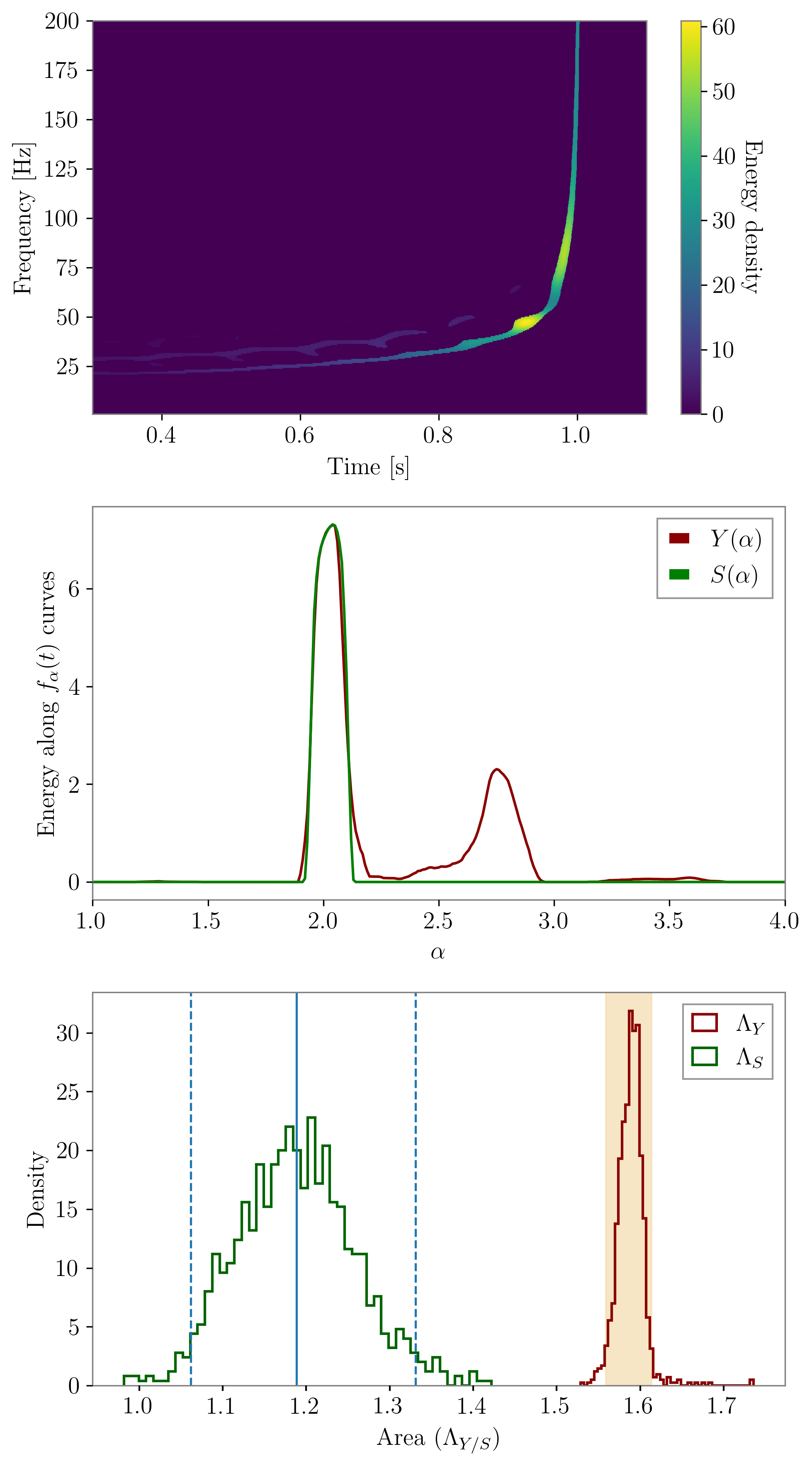} 
    \caption{\emph{Results for eccentric BBH injection analysis recovery with quasi-circular model.} the top panel shows the time-frequency map of the injected signal generated using synchroextracting transformation, the middle panel shows the comparison between the injected signal $Y(\alpha)$ and the template waveform $S(\alpha)$ corresponds to the maximum likelihood sample, and the bottom panel shows the comparison between the $\Lambda_S$ and $\Lambda_Y$ distributions. The vertical dashed lines indicate the 95\% intervals of the $\Lambda_S$ distribution, and the shaded band represents the 95\% intervals $\Lambda_Y$ distribution.}
    \label{fig:e}

\end{figure}

As a case study, we therefore evaluate the ability of our proposed consistency test to identify missing physics within GR by examining the impact of neglecting eccentricity in the search waveform model. We consider a numerical relativity (NR) waveform of an eccentric BBH system provided in the Simulating eXtreme Spacetimes (SXS) catalog. We specifically inject the \texttt{SXS:BBH:1360} waveform, which was simulated for a non-spinning equal mass eccentric binary \cite{Mroue:2013xna,Boyle:2019kee}. As the eccentricity evolves with time, we report the reference eccentricity \mbox{$e_{\rm ref}\sim 0.3636$} as measured at a reference orbital frequency of \mbox{$Mf_0=0.0153$}, and the reference mean anomaly \mbox{$l_{\rm ref} \sim 113.2^\circ$}. In our analysis, we scale the simulation for a total mass of $60\msun$, which is suitable for observing with 2nd generation detectors. So the reference eccentricity in this example corresponds to $f_0 \approx 50 \Hz$.

As prescribed in our methodology, we first perform the Bayesian analysis using a quasi-circular model \phXP and then we perform the consistency between the injection and the recovery model. We summarize the results in Fig.~\ref{fig:e}. The top-panel shows the time-frequency map of the injection, which clearly shows excitation of higher harmonics due to eccentricity. We note that the emission of this higher-harmonics are not due to higher-multipole moments, rather solely quadrupolar one. For a quasi-circular binary, the next-to-leading order harmonics is octupolar (m=3) one, for which the GW frequency is three times of the orbital frequency. Interestingly, for eccentric injection, the dominant signal power found at $\alpha \sim 2$, but the contribution of higher-harmonics spreaded  $\alpha\sim 2.5\text{---}3$ and peaking at $\alpha\sim2.75$. The previous method~\cite{Roy:2022teu} would fail to detect such higher harmonics contribution because it strictly look for signal power at integer values $\alpha$.

The bottom panel of Fig.~\ref{fig:e} shows the distributions for data $\Lambda_Y$ and the search waveforms $\Lambda_S$. The distributions are well separated and the average distance statistic value is $\avg{\DSY}\sim 5.84$, which is significantly above our reference threshold, as shown in Fig.~\ref{fig:d}. This implies a strong evidence of missing physics in \phXP model  to recover an eccentric signal.

\section{Analyzing GW190814 signal}
\label{sec:level4}

We finally demonstrate the sensitivity of our consistency test framework for identifying missing physics with real data. We apply this method to the GW190814 event~\cite{LIGOScientific:2020zkf}, assuming that the analyzing template waveform does not include higher-order modes. GW190814 is one of the exceptional events detected during the first half of the third observing run by the LVK Collaboration~\cite{LIGOScientific:2020zkf}. It was inferred that this signal was produced by the merger of a highly asymmetric binary with a mass ratio of approximately 9:1. This allowed us to find strong evidence of higher-order mode radiation in the GW. We investigate whether our method can indicate that a purely quadrupolar waveform is insufficient to represent the GW190814 signal. If so, what is the significance of the missing physics?

\begin{figure}[t!]
	\centering
		\includegraphics[width=0.95\linewidth]{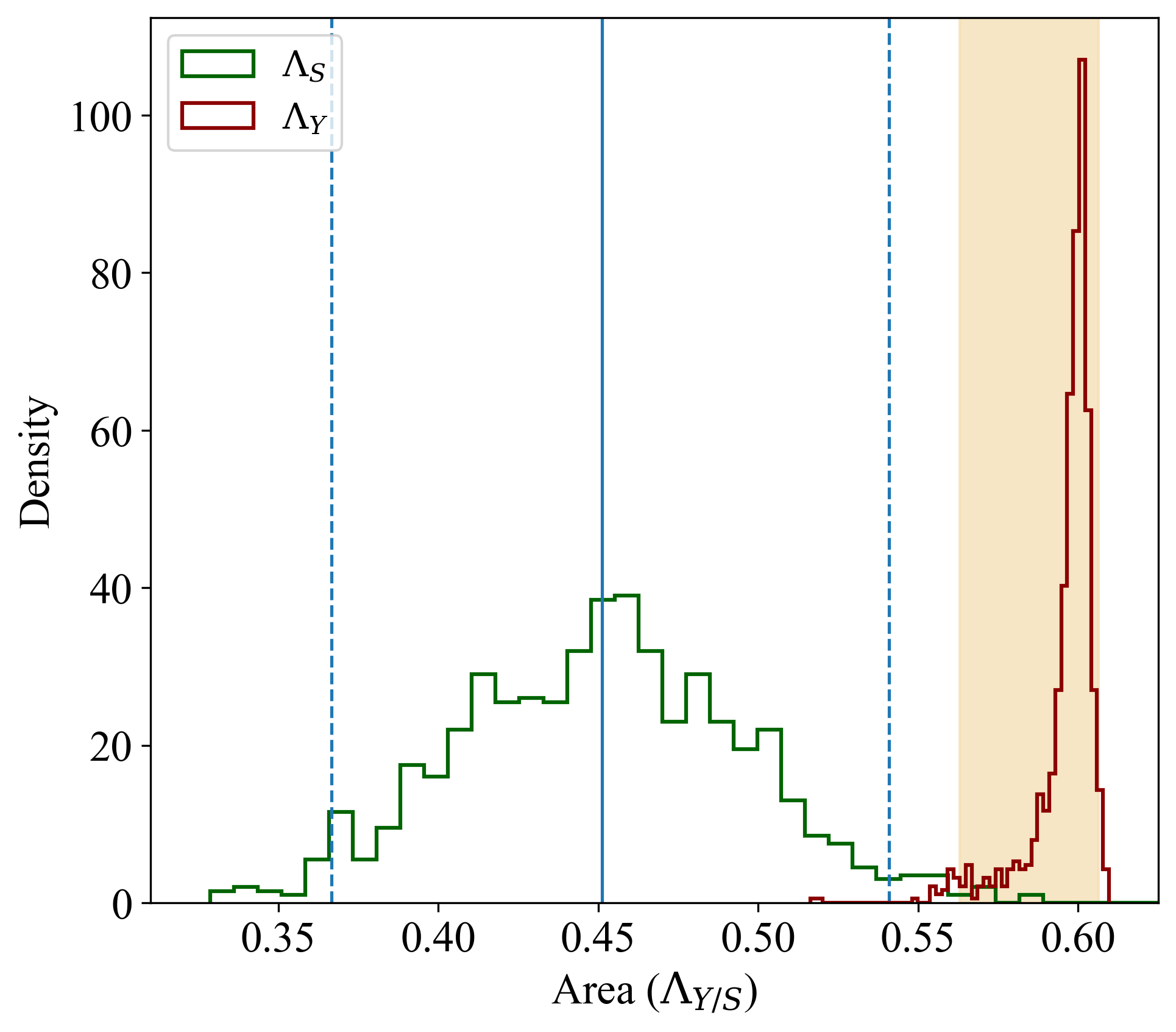} 
	\caption{Illustration of GW190814 results when the signal is analyzed with the \phXP waveform model. The distributions of $\Lambda_S$ and $\Lambda_Y${\textemdash}defined as the areas under the $S(\alpha)$ and $Y(\alpha)$ curves, respectively{\textemdash}are shown in green and maroon. The blue dashed lines represent the 95\% credible interval for the $\Lambda_S$ distribution, with the corresponding median indicated by a solid vertical line. The light maroon shaded region denotes the 95\% credible interval of the $\Lambda_Y$ distribution.}
	\label{fig:xp}
\end{figure}

To analyze the GW190814 signal, we use the strain data released with the GWTC-2.1 catalog by LVK~\cite{LIGOScientific:2021usb, KAGRA:2023pio}. In order to assess the effect of neglecting higher-order modes, we perform a parameter estimation using the \phXP model that incorporates only the dominant quadrupole mode. Using the \phXP posterior samples, we perform our new consistency test as described in Sec.~\ref{sec:level2b} and estimate the $\Lambda_Y$ and $\Lambda_S$ distributions. Fig.~\ref{fig:xp} shows the comparison $\Lambda_Y$ and $\Lambda_S$ distributions, 95\% intervals of these distributions are well separated. We also compute the distance between these two distributions $\avg{\DSY} \sim 3.24$ as highlighted in Fig.~\ref{fig:d}, which is above the reference threshold $3\sigma$ level. This implies that the quadrupolar waveform model \phXP is inadequate to represent the GW190814 signal.

Now the question is does this missing physics generated due to neglecting the higher-order modes in search waveforms?
We perform the consistency test using the publicly released \phXPHM posteriors~\cite{KAGRA:2023pio}. We found the distance statistic value $ \avg{\DSY} \sim 2.14 $, which is below the reference threshold and within the noise background. Therefore, we conclude that the deviation is generated due to neglecting the higher-order modes.

\section{Discussion}
\label{sec:level5}
In this study, we have introduced a new consistency test to determine the level of agreement between the observed signal and our theoretical waveform model. Deviations between the observed data and the search waveform model can manifest as changes in the distribution of pixel energy across the time-frequency plane. Our proposed method looks for consistency in the time-frequency plane by stacking the pixel energies along the time-frequency trajectories determined by the posterior waveforms. We find that this consistency test is efficient not only in detecting deviations from GR but also in identifying missing physics within GR itself. We summarize all results in Fig.\ref{fig:d} and Table\ref{tab:table1}.

To quantify the level of deviation from the search waveform model, our method yields the normalized deviation value \DSY, as defined in Eq.~\eqref{eq:distance}. According to our definition, the value of \DSY is always positive. As a general concern, a false deviation could be mimicked by the effect of instrumental noise. To interpret the value of \DSY, we have performed an extensive injection study assuming that the injection and recovered waveform belong to the same model. We find all such injections to have a significance level $\DSY < 2.4$. We therefore set the $3\sigma$ level as a reference threshold to assess whether the deviation is genuine rather than triggered by noise.

We have demonstrated the detectability of deviations for two types of injection studies: a GW170608-like massive graviton injection with graviton mass \Mgraviton as an example of a beyond-GR scenario, and a case study of missing physics due to unmodeled eccentricity. We have performed these injection studies assuming the designed sensitivity of Advanced LIGO and Virgo. Our new method is highly efficient in both cases; we find the distance statistic $\avg{\DSY}\sim 3.8$ and $\avg{\DSY}\sim 5.8$, respectively. This clearly indicates that such levels of deviation cannot be attributed to noise fluctuations.

For a quasi-circular binary, the next-to-leading-order harmonic is the octupolar $(m=3)$ mode, for which the GW frequency is three times the orbital frequency. The previous study by Roy et al.~\cite{Roy:2022teu} developed a method for detecting higher harmonics using the time–frequency trajectory, which looks for signal power at $\alpha = 3$ to identify the presence of octupolar contributions. For an eccentric binary, the emission of higher harmonics is not due to higher multipole moments but rather solely quadrupolar radiation. Interestingly, the dominant signal power is found at $\alpha \sim 2$, while the contribution of higher harmonics peaks around $\alpha \sim 2.75$. Since the previous method strictly looks for signal power at integer values of $\alpha$, it would fail to detect such higher–harmonic contributions driven by eccentricity. Another recent study employed a similar strategy using time–frequency trajectories for detecting higher harmonics in eccentric signals with an effective chirp–mass model~\cite{Hegde:2023yoz, Bose:2021pcw}. In contrast, our new method generalizes the strategy of the time–frequency trajectory to detect higher harmonics not only from quasi-circular binaries but also for eccentric cases.

We finally analyzed the GW190814 event to identify the presence of spherical higher harmonics, not eccentric ones. We first consider \phXP as the search waveform that only incorporates the quadrupolar mode. We found the distance statistic $\avg{\DSY}\sim 3.24$, which is above our reference threshold of noise, i.e., our chosen model is unable to represent the GW190814 signal. We then considered \phXPHM for the search model and found $\avg{\DSY}\sim 2.14$. This clearly indicates the strong evidence of higher harmonics in the GW190814 signal. We also performed this new consistency test for the GW190412 signal using these two models, and found $\avg{\DSY} \sim 1.42 $ and $\avg{\DSY} \sim 1.06$ for \phXP and \phXPHM models, respectively. Although the distance statistic values for both cases are within the noise level, the model with higher harmonics provides a better fit to the observed data.

We note that existing consistency methods either look for a coherent residual signal in the data after subtracting the best-fit posterior waveform or compute the overlap between the reconstructed signal and the posterior waveforms. Both of these methods rely on wavelet transformations. As the number of required wavelets to represent a signal increases as the signal duration becomes longer, even when the SNR remains the same, the efficiency of these methods drops for longer signals. In contrast, our time-frequency trajectory-based consistency test remains sensitive for detecting deviations in longer signals.

\begin{acknowledgments}
We are grateful to Haris M. K. and Justin Janquart for useful comments on our draft. This work was supported by the National Natural Science Foundation of China (NSFC), Grant No.~W2442005, 12250610185 and 12261131497, and the Natural Science Foundation of Shanghai, Grant No.~22ZR1403400. D.~D. also acknowledges support from the China Scholarship Council (CSC), Grant No. 2022GXZ005434. S.R. is supported by the Fonds de la Recherche Scientifique - FNRS (Belgium). D. D. thanks Universit{\'e} catholique de Louvain for hospitality and support. The manuscript is based upon work supported by NSF's LIGO Laboratory, which is a major facility fully funded by the National Science Foundation (NSF), as well as the Science and Technology Facilities Council (STFC) of the United Kingdom, the Max-Planck-Society (MPS), and the State of Niedersachsen/Germany for support of the construction of Advanced LIGO and construction and operation of the GEO600 detector. Additional support for Advanced LIGO was provided by the Australian Research Council. Virgo is funded through the European Gravitational Observatory (EGO), by the French Centre National de Recherche Scientifique (CNRS), the Italian Istituto Nazionale di Fisica Nucleare (INFN) and the Dutch Nikhef, with contributions by institutions from Belgium, Germany, Greece, Hungary, Ireland, Japan, Monaco, Poland, Portugal, Spain. KAGRA is supported by the Ministry of Education, Culture, Sports, Science and Technology (MEXT), Japan Society for the Promotion of Science (JSPS) in Japan; National Research Foundation (NRF) and the Ministry of Science and ICT (MSIT) in Korea; Academia Sinica (AS) and National Science and Technology Council (NSTC) in Taiwan. We have used \numpy~\cite{Harris:2020xlr}, \scipy~\cite{Virtanen:2019joe}, \matplotlib~\cite{Hunter:2007ouj} for analyses and preparing the figures in the manuscript.
\end{acknowledgments}


\appendix

\begin{figure*}[ht]
\centering
\includegraphics[width=0.9\linewidth]{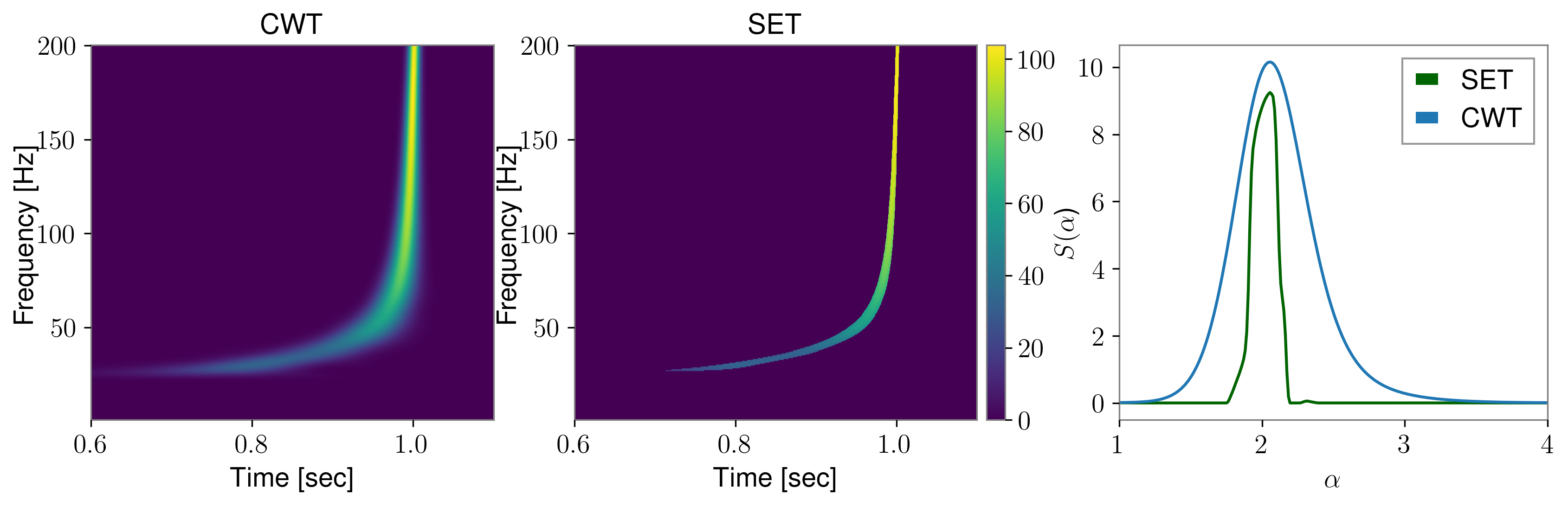} 
\caption{Comparison between the CWT (left panel) and SET (middle panel) of a $(30, 30)\msun$ BBH waveform generated using the \phXP model, along with their respective $S(\alpha)$ curves shown in the right panel. SET results in better TF resolution, which translates to better resolution in $\alpha$. Good resolution in $\alpha$ is very important to make the statistics work efficiently for our project.}
	\label{fig:0}
\end{figure*}
\section{Synchroextraction Transform (SET)}
\label{sec:A}

Time–frequency (TF) spectrograms provide a direct visual representation of the frequency evolution of compact binary systems. However, most commonly used TF methods suffer from time–frequency resolution limits, which can spread signal energy into neighboring pixels and lead to blurred structures in the TF plane. This leakage arises from the trade-off between time and frequency resolution and from the choice of windowing, which affects time–frequency concentration. We address this issue using the recently developed synchroextracting transform (SET), based on reassignment techniques~\cite{yu2017synchroextracting, pham2017high}. The resulting high-resolution TF representation concentrates signal energy into fewer pixels. This sharper representation helps us detect subtle changes in the frequency evolution that may arise in beyond-GR scenarios compared to the GR prediction.

The primary concept of SET is to retain/extract the TF information of continuous wavelet transformation (CWT) most related to time-varying features of the GW signal, hence resulting in the removal of most of the smeared TF energy. For an accurate localization of the signal energy in the TF plane, SET involves the estimation of the instantaneous frequency, $\hat{f}(t,f)$. This approach to calculate $\hat{f}(t,f)$ is very efficient even in the presence of noise. 
\begin{equation} \label{eq:if}
    \hat{f} (t,f) = f + \frac{i}{2 \pi} \frac{\tilde{W}^{w'} (t ,f )}{\Tilde{W}^w (t,f)},
\end{equation}
where, $\Tilde{W}^{w} (t,f )$ and $\Tilde{W}^{w'} (t,f )$ are the CWTs of the GW signals taken with a Gabor-Morlet wavelet ($w$) and time derivative of Gabor-Morlet wavelet ($w'$). For the given definition of instantaneous frequency in Eq.~(\ref{eq:if}), the SET can be written as 
\begin{align}
\label{eq:set}
  \mathrm{SET}[\Tilde{W}^w (t, &f)] 
    = \int  \Tilde{W}^w (t,f) \; \delta \left( f - \hat{f} (t,f)\right) df \nonumber \\
    &= \int  \Tilde{W}^w (t,f) \; \delta \left(- \frac{i}{2 \pi} \frac{\Tilde{W}^{w'} (t ,f )}{\Tilde{W}^w (t,f)}\right) df
\end{align}
The above formulation implies that SET is a post-processing method that keeps only the coefficients aligned with the instantaneous-frequency ridges. Thereby, the SET representation in general does not preserve the total energy of the original spectrogram, but suitable for identifying the instantaneous time-frequency trajectories.

Fig.~\ref{fig:0} shows the spectrograms corresponding to $\Tilde{W}^w (t,f)$ and SET$[\Tilde{W}^w (t,f)]$ for a $(30, 30) \textup{M}_\odot$ BBH system. The Advanced LIGO and Virgo power spectral densities are used to whiten the GW strain. We note that the (2,2) mode energy is highly localized in fewer pixels in the SET spectrogram. The $S(\alpha)$ vectors also show a much narrower peak at $\alpha \thickapprox 2$ for the SET compared to the peak for the CWT spectrogram.


\section{Accessing the impact of noise on consistency test}
\label{sec:B}
Even when the true signal model matches our theoretical waveform, detector noise alone can produce apparent deviations. To assess the impact of instrumental noise on the sensitivity of our framework, we perform an extensive injection study. As described in Sec.~\ref{sec:level2c}, in order to isolate the effect of noise, we generate the injections using the \phXP model and recover them with the same model, keeping the binary parameters identical across all simulations. We simulate the injections in stationary Gaussian noise weighted by the Advanced LIGO and Virgo sensitivity curves. Applying our proposed consistency test, we obtain the distributions for the data, $\Lambda_Y$, and for the search waveforms, $\Lambda_S$. For each simulation, we construct the $\Lambda_S$ and $\Lambda_Y$ distributions using 1000 randomly selected posterior samples.

\begin{figure*}
\includegraphics[width=0.90\textwidth]{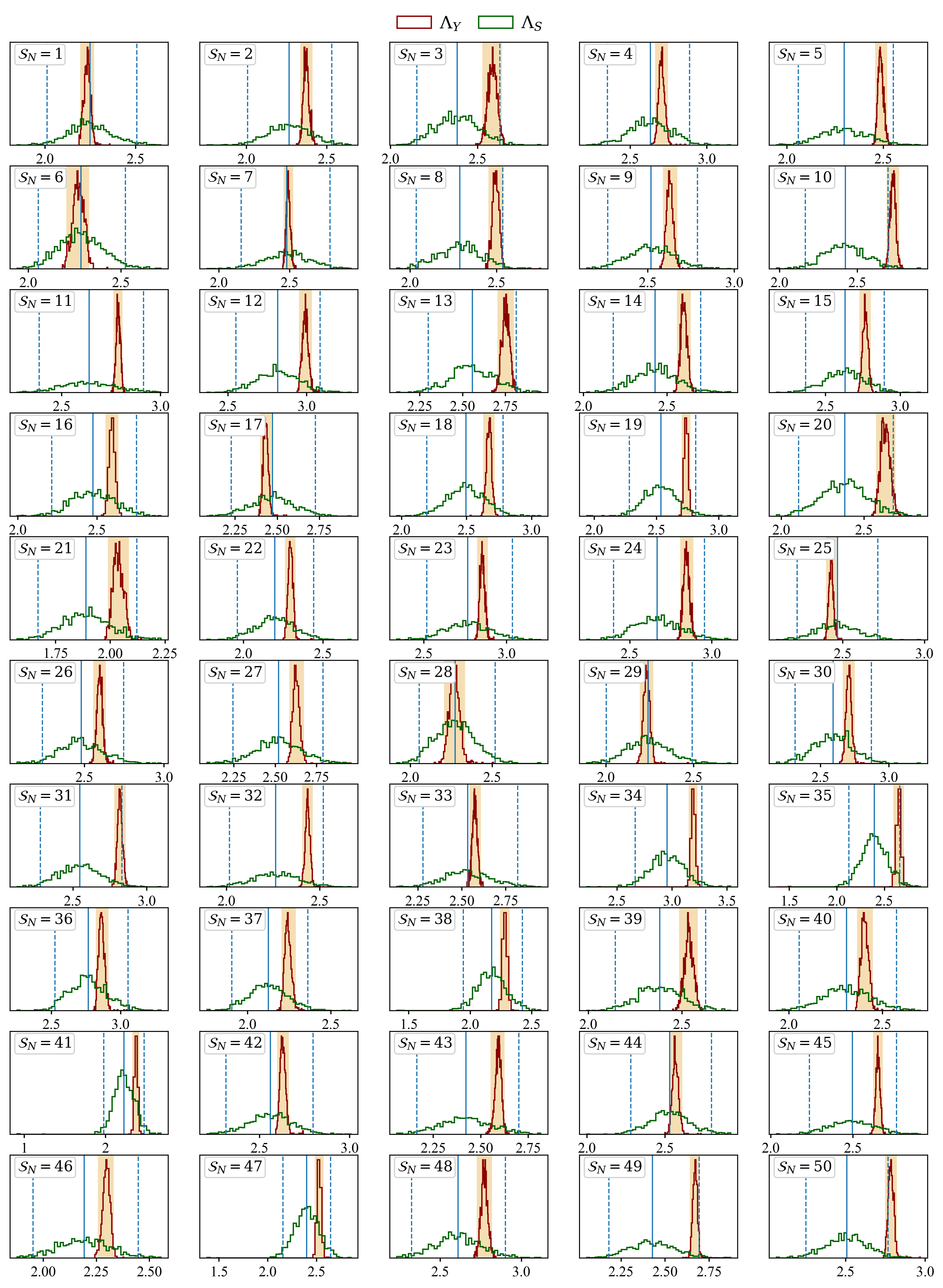}
\caption{Comparison between \sigmaS (green) and $\Lambda_Y$ (maroon) distributions is shown for simulated signals in noise, where the injection and recovery waveforms are generated within the same model \phXP. The quantity $\mathcal{S}_N$ in each panel represents the simulation number. The light brown shaded region represents the 95\% interval of the $\Lambda_Y$ distribution. The solid vertical line and the dashed vertical lines denote the mean and 95\% intervals of $\Lambda_S$ distribution, respectively.}
\label{fig:n}
\end{figure*}

\begin{figure*}
\begin{center}
\includegraphics[width=0.90\textwidth]{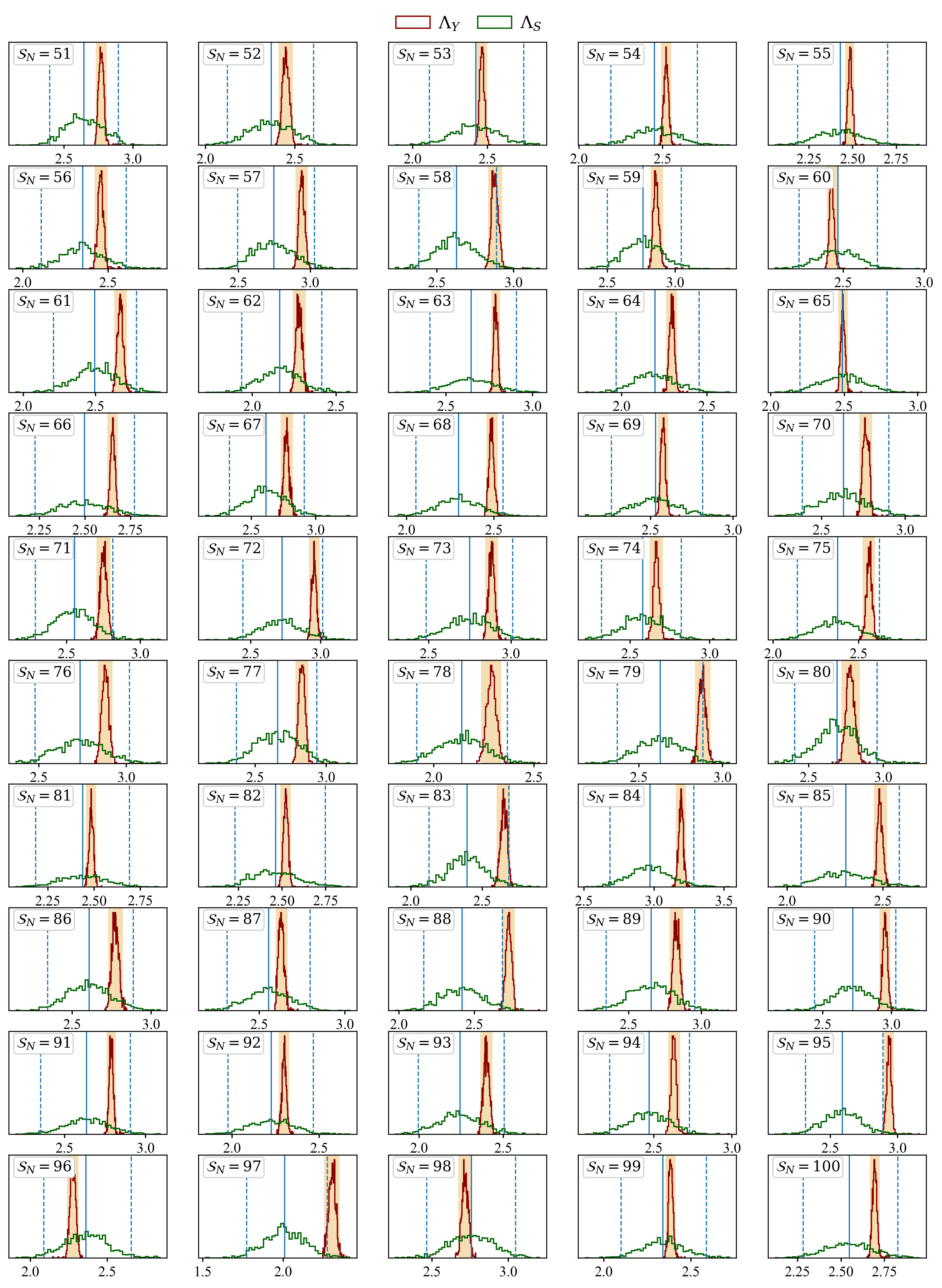}
\caption{Same as Fig.~\ref{fig:n}, but displaying the comparison between $\Lambda_S$ and $\Lambda_Y$ distributions for next 50 simulations ($\mathcal{S}_N = 51$ to $100$). 
}
\label{fig:n2}
\end{center}
\end{figure*}

Fig.\ref{fig:n} and Fig.\ref{fig:n2} show the comparison between the $\Lambda_Y$ and $\Lambda_S$ distributions. The tag $\mathcal{S}_N$ denotes the serial index of the simulations used in the analysis. Two common features are observed: compared to the $\Lambda_S$ distributions, the $\Lambda_Y$ distributions are significantly narrower and mostly shifted toward larger values. The first feature is analogous to the well-known behaviour of the matched-filter SNR versus the optimal SNR across posterior samples. In our case, the time–frequency tracks change only slightly across posterior samples, and therefore the $Y(\alpha)$ curve remains nearly unchanged, whereas the $S(\alpha)$ curve varies from sample to sample due to dependence on extrinsic parameters such as distance, sky location, polarization angle, and reference time and phase. The second feature arises from the fact that $Y(\alpha)$ curves are constructed from data that contain signal plus noise. The additional contribution of noise power in the time–frequency pixels shifts the $\Lambda_Y$ distributions toward larger values relative to $\Lambda_S$. For zero-noise injections, however, the $\Lambda_Y$ distribution is consistent with the median of the $\Lambda_S$ distribution.

As shown in Fig.~\ref{fig:d}, we compute the normalized deviation between the $\Lambda_Y$ and $\Lambda_S$ distributions. The distance statistic satisfies $\avg{\DSY} < 2.4$ for all injections. We tentatively set \(\DSY = 3\) as a reference threshold for detecting deviations.

\clearpage
\bibliography{references}

\end{document}